\documentclass[12pt]{article}
\pdfoutput=1
\usepackage{amsmath}
\usepackage{amssymb} 
\usepackage{bbm} 
\usepackage[small]{caption2} 
\usepackage{fleqn} 
\usepackage{graphicx} 
\usepackage{mathrsfs}   
\usepackage[small,loose]{subfigure}  
\usepackage{cite} 
\usepackage{color}
\usepackage{xspace}
\usepackage{graphicx}
\usepackage{color}

\graphicspath{{./Abbildungen/}}
\newcommand{\eVdist}{\kern-0.06em}

\newcommand{\gev}{\:\text{Ge\eVdist V}}

\newcommand{\Z}[1]{\ensuremath{\mathbbm{Z}_{#1}}} 

\newcommand{\be}{\begin{equation}}
\newcommand{\ee}{\end{equation}}
\newcommand{\bea}{\begin{eqnarray}}
\newcommand{\eea}{\end{eqnarray}}

\newcommand{\SARAH}{{\tt SARAH}\xspace}
\newcommand{\SPheno}{{\tt SPheno}\xspace}

\allowdisplaybreaks[1]

\begin{document}

\begin{titlepage}

\vspace*{-3.0cm}
\begin{flushright}
CERN-PH-TH/2012-301 \\
Bonn-TH-2012-29 \\
DESY 12-203
\end{flushright}

\begin{center}
{\Large\bf
  Enhanced diphoton rates at Fermi and the LHC
}

\vspace{1cm}

\textbf{
Kai Schmidt-Hoberg$^a$,
Florian Staub$^b$,
Martin Wolfgang Winkler$^c$
}
\\[5mm]
\textit{$^a$\small
Theory Division, CERN, 1211 Geneva 23, Switzerland
}
\\[5mm]
\textit{$^b$\small
Bethe Center for Theoretical Physics \& Physikalisches Institut der 
Universit\"at Bonn, \\
Nu{\ss}allee 12, 53115 Bonn, Germany
}
\\[5mm]
\textit{$^c$\small
Deutsches Elektronen-Synchrotron DESY, \\
Notkestra{\ss}e 85, D-22607 Hamburg, Germany
}
\end{center}

\vspace{1cm}

\begin{abstract}
 We show that within MSSM singlet extensions the experimental hints beyond the standard model from the Fermi LAT telescope as well
as from the LHC can be explained simultaneously while being consistent with all experimental constraints. In particular we present
an example point which features a $\sim 130 \gev$ lightest neutralino with an annihilation cross section into photons consistent with the
indication from the Fermi satellite with simultaneously the right relic abundance, a continuum photon spectrum consistent with observation,
direct detection cross section below the experimental limits, electroweak observables consistent with experiment and a
$125 \gev$ light Higgs boson with a slightly enhanced $h \rightarrow \gamma \gamma$ rate.  
 \end{abstract}

\end{titlepage}

\section{Introduction}

2012 has been a very intriguing year regarding hints for new physics, both at the LHC and the Fermi large area telescope.
While the data from ATLAS~\cite{ATLAS:2012} and CMS~\cite{CMS:2012} feature a new bosonic state with mass $m \sim 125 \gev$ which is consistent with the expectation from a standard model (SM) Higgs, both experiments see indications for an excess in the diphoton channel. 
This potential enhancement in the diphoton rate has attracted much attention recently, see e.g.~\cite{Ellwanger:2011aa,Cao:2012fz,Barger:2012ky,Carena:2012xa,Alves:2012ez,Bonne:2012im,Bellazzini:2012mh,Buckley:2012em,An:2012vp,Cohen:2012wg,Alves:2012yp,Joglekar:2012hb,Haba:2012zt,Almeida:2012bq,Benbrik:2012rm,ArkaniHamed:2012kq,SchmidtHoberg:2012yy,Chala:2012af}.
In contrast the diboson decays into $WW^*$ and $ZZ^*$ seem to be in accord with the SM expectation, which make explanations
of the enhanced diphoton rate due to an increased partial decay width particularly appealing.

A similarly exciting topic this year have been the hints for a $\gamma$-ray line in the Fermi LAT data as reported in~\cite{Bringmann:2012vr,Weniger:2012tx}. $\gamma$ ray lines are considered the smoking gun of annihilating dark matter, as astrophysical processes able to induce line-like features are very rare (see however~\cite{Profumo:2012tr,Aharonian:2012cs}). Intriguingly, the morphology of the excess is consistent with the expected distribution of dark matter up to a small offset from the galactic center~\cite{Su:2012ft,Bringmann:2012ez}.\footnote{Recent numerical simulations indicate that such a small offset can indeed be realized in realistic models of galaxy formation~\cite{Kuhlen:2012qw}.}
Currently the data are being re-analysed by the Fermi collaboration, and there seems to be an indication of a line-like feature at a slightly higher energy of 135~GeV~\cite{FermiTalk}, where the shift results from a reprocessing of the data. The statistical significance of the excess found by the Fermi collaboration is, however, not as high as claimed in \cite{Weniger:2012tx,Tempel:2012ey,Su:2012ft}, although this also depends on the target region considered.
A line feature has also appeared in the $\gamma$ ray data of the earth limb, raising some concerns about an instrumental effect~\cite{Hektor:2012ev,Finkbeiner:2012ez,FermiTalk}. Radio telescopes might help to confirm or rule out the dark matter interpretation of the line soon \cite{Laha:2012fg}. While the origin of the $\gamma$-ray line from the galactic center still has to be clarified, a noticeable amount of theoretical interest has been triggered~\cite{Chalons:2012hf,Chalons:2011ia,Ibarra:2012dw,Dudas:2012pb,Cline:2012nw,Choi:2012ap,Kyae:2012vi,Lee:2012bq,Rajaraman:2012db,Buckley:2012ws,Chu:2012qy,Das:2012ys,Kang:2012bq,Weiner:2012cb,Buchmuller:2012rc,Frandsen:2012db,D'Eramo:2012rr}. If interpreted in terms of dark matter, the $\gamma$-ray line requires a rather large annihilation cross section into photons, $\langle\sigma v\rangle_{\gamma\gamma}\sim 10^{-27}\:\text{cm}^3\:\text{s}^{-1}$, if one assumes an Einasto profile~\cite{Weniger:2012tx}. Such a large cross section was found to be very difficult to accommodate in particle physics models, especially as the Fermi data are consistent with pure background at lower energies, i.e.\ competing annihilation channels must be sufficiently suppressed~\cite{Buckley:2012ws,Buchmuller:2012rc,Cohen:2012me,Cholis:2012fb}.

While one may tackle each of these experimental anomalies individually, it is intriguing to speculate about a possible common origin. 
In this article we show that the signals from both Fermi and the LHC can be explained simultaneously within singlet extensions of the MSSM, while being consistent with
all experimental constraints. Singlet extensions are particularly interesting given the observed value for the Higgs mass because
the electroweak fine-tuning can be substantially alleviated in these models.
A somewhat generalised version of the NMSSM, the GNMSSM~\cite{Ross:2011xv}, which is based on a discrete $R$ symmetry \cite{Lee:2010gv, Lee:2011dya}, was found to be particularly promising in this context~\cite{Ross:2012nr}.
In \cite{SchmidtHoberg:2012yy} it was found that in this setup the coupling of the CP even neutral light Higgs to light charginos can be strongly enhanced, leading to a sizeable increase in the $h \rightarrow \gamma \gamma$ rate.
Interestingly the same coupling can lead to a rather large neutralino annihilation cross section into photons as indicated by the Fermi data,
while being compatible with bounds from direct detection, electroweak precision observables, the continuum photon spectrum 
and with a $125\gev$ Higgs with an enhanced diphoton rate.

This article is organised as follows: In the next section we will briefly review some aspects of the GNMSSM. In section~\ref{sec:annihilation} we will then
discuss neutralino annihilation within this framework, before we come to constraints arising from the requirement of the correct relic abundance and the 
continuum photon spectrum in section~\ref{sec:continuum}. Constraints from direct and indirect detection experiments are analysed in section~\ref{sec:dd}.
Section~\ref{sec:spheno} is then devoted to a thorough numerical study of a benchmark point,
while section~\ref{sec:conclusions} contains our summary. Some useful information about the GNMSSM is collected in the appendix.

\section{The GNMSSM}
\label{sec:gnmssm}

As a framework, we consider the GNMSSM, a generalised version of the NMSSM, which has a superpotential of the form
\begin{eqnarray}
 \mathcal{W} &=& \mathcal{W}_\text{Yukawa}  + \frac{1}{3}\kappa S^3+
(\mu + \lambda S) H_u H_d +  \frac{1}{2} \mu_s S^2  \;.
\end{eqnarray}
Here $\mathcal{W}_\text{Yukawa}$ are the MSSM superpotential terms generating the usual Yukawa couplings and we used the freedom to shift the singlet $S$ to set a potential linear term in $S$ to zero. 
This superpotential has additional explicit mass terms $\mu$ and $\mu_s$ which are not present in the $\Z{3}$ symmetric NMSSM which is usually considered (for reviews of the NMSSM see e.g.~\cite{Maniatis:2009re,Ellwanger:2009dp}). While the apparent un-naturalness of these additional
mass terms has prevented a larger community from studying the phenomenology of the GNMSSM,
it has recently been realised that exactly this structure naturally arises from an underlying $R$ symmetry as discussed in \cite{Lee:2011dya}.
The fact that this $R$ symmetry also eliminates the dangerous dimension four and five baryon- and lepton-number violating terms and avoids destabilising tadpoles and domain wall problems makes it a more promising starting point than the $\Z{3}$ symmetric NMSSM. 

The soft SUSY breaking terms associated with the extended Higgs sector of the GNMSSM are given by
\begin{align}
 V_\text{soft} 
   &=  m_s^2 |s|^2 + m_{h_u}^2 |h_u|^2+ m_{h_d}^2 |h_d|^2 \nonumber \\
   &+ \left(b\mu \, h_u h_d + \lambda A_\lambda s h_u h_d + \frac{1}{3}\kappa A_\kappa s^3 + \frac{1}{2} b_s s^2  + \xi_s s + h.c.\right) \;.
\label{soft}
\end{align}
The resulting mass matrices as well as the relevant couplings for our discussion are given in appendix~\ref{sec:massmatrices} and \ref{sec:couplings}. 
The field content is the same as in the NMSSM.
In comparison with the MSSM there is an additional singlet fermion which mixes into the neutralino sector and an additional complex scalar which
mixes into the Higgs sector. For more details on the model, see~\cite{Ross:2011xv,Ross:2012nr}.

\begin{figure}[hbt]
\centering
\includegraphics[width=14cm]{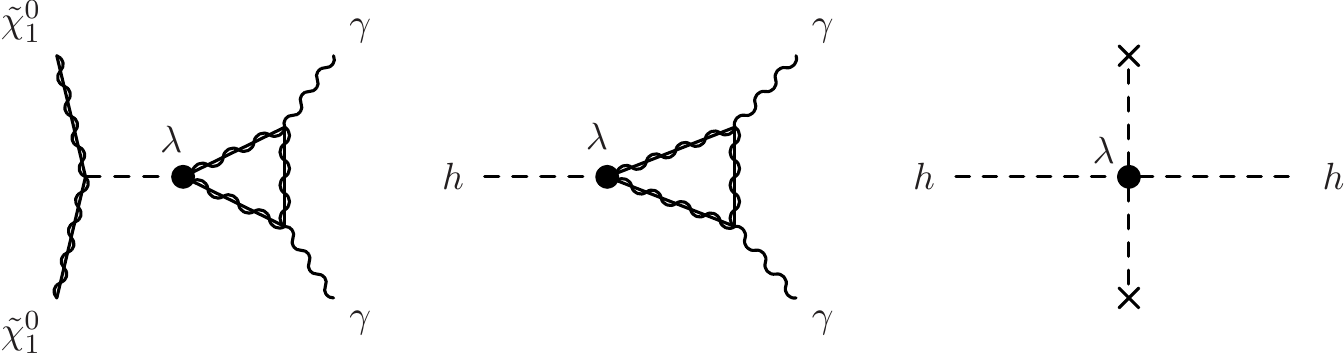}\hspace{2cm}
\caption{In the GNMSSM the $\lambda$-coupling drives the annihilation of neutralinos into photons (left), enhances the Higgs decay rate into photons (middle) and provides an additional contribution to the Higgs boson mass (right).}
\label{fig:benefit}
\end{figure}

The Higgs sector of the GNMSSM has been discussed in \cite{Ross:2011xv,Ross:2012nr,SchmidtHoberg:2012yy}. As in the NMSSM there
is an additional tree-level contribution to the lightest Higgs mass, which is large for small $\tan\beta$ and large $\lambda$,
allowing to evade the upper bound of $m_h < M_Z$ present in the MSSM.
The fact that radiative corrections due to top/stop loops are no longer needed drastically reduces the electroweak fine-tuning in MSSM singlet extensions~\cite{BasteroGil:2000bw,Dermisek:2005gg,Dermisek:2006py,Dermisek:2007yt,Ellwanger:2011mu} for large values of $\lambda$.
Given the Higgs mass of 125~GeV the GNMSSM has been shown to be particularly interesting in this context \cite{Ross:2011xv,Ross:2012nr}. 
One additional advantage of the GNMSSM is the fact that for small $v_s$ no tuning in the Higgs mass matrix is required to avoid large doublet-singlet mixing\footnote{
In the NMSSM it is common practice to tune the parameter $A_\lambda$ such that the off-diagonal entry in the mass matrix is close to zero,
avoiding doublet-singlet mixing and allowing for a correspondingly larger Higgs mass.}.
It has also been argued that allowing even larger values of $\lambda$, relaxing the condition that it remain perturbative up to the GUT scale, leads
to an additional reduction in the fine-tuning \cite{Hall:2011aa}.

As discussed in \cite{SchmidtHoberg:2012yy} scenarios with large $\lambda$ may also accommodate an enhanced diphoton rate as observed by ATLAS and CMS: in addition to the dominant $W$ and top loops present in the SM, new sizeable contributions from charged Higgs and chargino loops can arise, if these states are light. While a light charged Higgs is generically challenged by the measurement of $b\to s\gamma$, a light chargino is perfectly viable and an interesting possibility.
In the next section we will show that large $\lambda$ and light charginos can also lead to a large annihilation cross section into photons, see also figure~\ref{fig:benefit}.

\section{Neutralino annihilation into photons}
\label{sec:annihilation}

The mass matrix as well as the relevant couplings of the GNMSSM neutralino sector can be found in the appendix.
\begin{figure}[hbt]
\centering
\includegraphics[width=5cm]{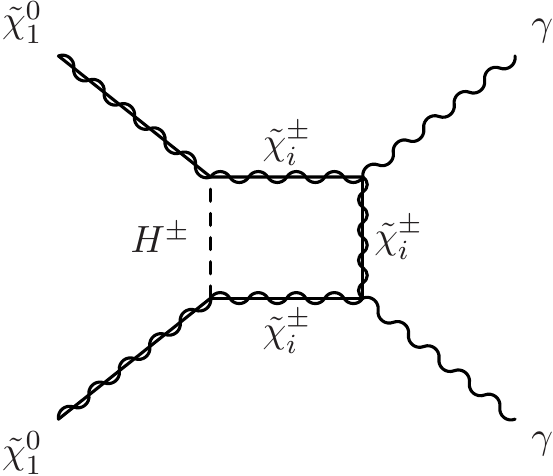}\hspace{2cm}
\includegraphics[width=5cm]{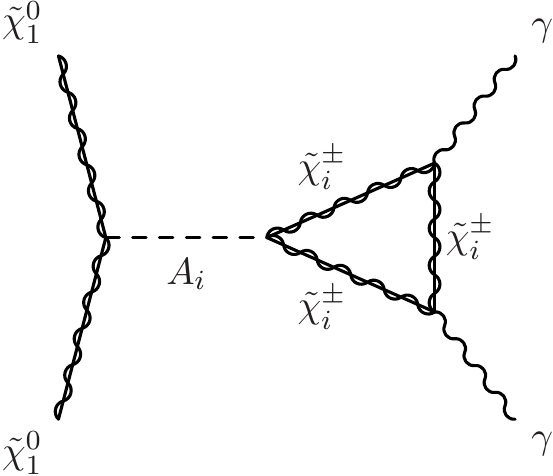}
\caption{Relevant Feynman diagrams for annihilation of the lightest neutralino into two photons within the GNMSSM.}
\label{fig:annihilation}
\end{figure}
Compared to the MSSM, the GNMSSM offers two alternative possibilities to achieve a large annihilation cross section $\langle\sigma v\rangle_{\gamma\gamma}$. The corresponding Feynman diagrams are shown in figure~\ref{fig:annihilation}. In case $\tilde{\chi}^0_1$ carries a sizeable singlino fraction, the chargino/ charged Higgs loop on the left experiences a drastic enhancement compared to the MSSM through the $\lambda$-coupling. However, even for $\lambda>1$, a cross section large enough to explain the Fermi line requires not only the charginos but also the charged Higgs bosons to be very light. This typically causes problems with flavour observables, in particular $b\rightarrow s \gamma$. Therefore, we will concentrate on the second possibility in the following, shown in the right panel of figure~\ref{fig:annihilation}. The diagram contains a pseudoscalar Higgs $A_i$ in the s-channel, i.e.\ the cross section can be enhanced in the vicinity of the pseudoscalar resonance (see also \cite{Das:2012ys}).

Analytically, one finds (see e.g.~\cite{Bergstrom:1997fh,Bern:1997ng})
\begin{equation}
 \langle \sigma v\rangle_{\gamma\gamma} = \frac{\alpha^2 m^2_{\tilde{\chi}^0_1}}{16 \pi^3} \left|\; 
\tilde{\mathcal{A}}\; \right|^{2}\;.  
\label{eq:sigmav} 
\end{equation} 
with
\begin{equation}
 \tilde{\mathcal{A}}_{A_1}  = 
\sum_{\tilde{\chi}^\pm_i} \frac{m_{\tilde{\chi}^\pm_i}}{2 \, m_{\tilde{\chi}^0_1}} \; 
   \frac{g_{\tilde{\chi}^0_1  \tilde{\chi}^0_1  A_1} \, g_{\tilde{\chi}^\pm_i \tilde{\chi}^\pm_i A_1}}{(4\,m^2_{\tilde{\chi}^0_1}
   -m^2_{A_1})}\;\arctan^2\left[\left(\frac{m^2_{\tilde{\chi}^\pm_i}}{m^2_{\tilde{\chi}^0_1}}-1\right)^{-1/2}\right]\;.  
\end{equation}
Here $g_{\tilde{\chi}^0_1  \tilde{\chi}^0_1  A_1} \equiv g_{\tilde{\chi}^0_1  \tilde{\chi}^0_1  A_1}^R-g_{\tilde{\chi}^0_1  \tilde{\chi}^0_1  A_1}^L$ and 
$g_{\tilde{\chi}^\pm_i \tilde{\chi}^\pm_i A_1} \equiv g_{\tilde{\chi}^\pm_i \tilde{\chi}^\pm_i A_1}^L-g_{\tilde{\chi}^\pm_i \tilde{\chi}^\pm_i A_1}^R$.
The general form of the trilinear couplings $g_{\tilde{\chi}^0_1\tilde{\chi}^0_1 A_1}^{L,R}$ and  $g_{\tilde{\chi}^\pm_i \tilde{\chi}^\pm A_1}^{L,R}$ is given in appendix~\ref{sec:couplings}.
Throughout this article we will assume $\lambda$ and $\kappa$ to be real.

An interesting limit arises if we assume that $\tilde{\chi}^0_1$ and $A_1$ are dominantly singlet-like. In this case, we obtain $g_{\tilde{\chi}^0_1  \tilde{\chi}^0_1  A_1}=\sqrt{2}\kappa$. Further, only the higgsino-like chargino couples to $A_1$ with $g_{\tilde{\chi}^\pm_1 \tilde{\chi}^\pm_1 A_1}=\lambda/\sqrt{2}$ for the lighter chargino being a pure higgsino. Setting as an example $m_{\tilde{\chi}^\pm_1}=1.5\,m_{\tilde{\chi}^0_1}$, we obtain in this limit
\begin{equation}
\langle \sigma v\rangle_{\gamma\gamma} \simeq (6\cdot10^{-28}\;\text{cm}^3\;\text{s}^{-1}) \cdot \lambda^2 \kappa^2 \left(\frac{(100\gev)^2}{4\,m^2_{\tilde{\chi}^0_1} -m^2_{A_1}}\right)^2 \left(\frac{m_{\tilde{\chi}^0_1}}{130\gev}\right)^2 \;.
\end{equation}
This shows that a cross section large enough to explain the Fermi line can indeed be realised if we allow for a mild tuning of $m_{A_1}$. In the above example, $m_{A_1}$ has to be within the range $m_{A_1}\simeq 240-280\gev$ (if we assume $\lambda,\kappa \lesssim 1$). The tuning is, however, substantially less severe than in the NMSSM case~\cite{Das:2012ys}. The reason is that in the NMSSM it is not possible to obtain a dominantly singlet like pseudoscalar and a singlino like LSP with the desired
masses simultaneously. Instead the lightest neutralino has to be predominantly bino like, leading to much smaller effective couplings.
Large enough annihilation cross sections can only be achieved very close to the pseudoscalar resonance, which in turn requires substantial tuning,
in particular, when the most recent XENON100 bounds~\cite{Aprile:2012nq} are taken into account, which appeared after~\cite{Das:2012ys} was published.

In addition to the $\gamma$ ray line from annihilation into two photons, there is a second line from the process $\tilde{\chi}^0_1 \tilde{\chi}^0_1 \rightarrow Z\gamma$ at a slightly lower energy
\begin{equation}
 E_\gamma = m_{\tilde{\chi}^0_1 }\, \left( 1- \frac{M_Z^2}{4\,m_{\tilde{\chi}^0_1 }^2} \right)\;.
\end{equation}
As has been shown in~\cite{Su:2012ft} such a double line structure even slightly improves the fit to the Fermi data compared to a single line.
The cross section $\langle \sigma v\rangle_{Z\gamma}$ can be calculated with the formulas presented in~\cite{Ullio:1997ke}. We find that for a pure higgsino $\langle \sigma v\rangle_{Z\gamma}/\langle \sigma v\rangle_{\gamma\gamma} \sim 0.6$. The wino, however, has a stronger coupling to $Z$ bosons than the higgsino. Therefore, chargino mixing tends to increase the relative importance of the $Z\gamma$-channel.

\section{Continuum photons and relic density}
\label{sec:continuum}

Even if dark matter annihilations induce a $\gamma$ ray line consistent with Fermi, it remains quite challenging not to overproduce continuum $\gamma$s by competing annihilation processes. Irrespective of whether dark matter is produced thermally or non-thermally, its present-day annihilation fraction into $\gamma$s must satisfy~\cite{Buchmuller:2012rc}
\begin{equation}
 \text{Br}_{\gamma\gamma}=\frac{\langle\sigma v\rangle_{\gamma\gamma}}{\langle\sigma v\rangle} \gtrsim 10^{-2} \;.
\end{equation}
For the winos and higgsinos of the MSSM, this fraction is in the range $\text{Br}_{\gamma\gamma}= \mathcal{O}(10^{-3})$. Therefore, the low energy $\gamma$ data disfavour an explanation of the Fermi line within the framework of the MSSM~\cite{Buchmuller:2012rc,Cohen:2012me}.

In the GNMSSM there is a simple possibility to suppress the continuum $\gamma$s. As in~\cite{Das:2012ys}, we take the lighter pseudoscalar $A_1$ to be an almost pure singlet. With the appropriate choice of $b\mu$ and $b_s$ it is simple to arrive at a situation where the singlet pseudoscalar remains light while the MSSM pseudoscalar becomes heavy and decouples. In this case, any tree level annihilation process into quarks through an intermediate $A_1$ is suppressed by the mixing angle between singlet and MSSM pseudoscalar. More specifically, for $\lambda\sim1$ the effective coupling of $A_1$ to photons becomes comparable to the tree-level coupling to bottom quarks for a doublet fraction at the level of $0.1\%$. This can easily be achieved for a sufficiently heavy $A_2$ without requiring any cancellations in the pseudoscalar mass matrix.

Competing annihilation processes into SM states which do not proceed through $A_1$ can be suppressed if  $\tilde{\chi}^0_1$ is dominantly singlino-like. A singlino-like LSP arises if the gauginos and higgsinos are sufficiently heavier than the singlino.
With a singlino-like $\tilde{\chi}^0_1$ and a singlet-like $A_1$ one naturally obtains $\text{Br}_{\gamma\gamma} \gg 10^{-2} $. 

Within the GNMSSM it is also possible to realise thermal dark matter and the $\gamma$ ray line simultaneously. Thermal production requires a total neutralino annihilation cross section $\langle\sigma v\rangle_\text{FO}\sim 2\cdot 10^{-26}\:\text{cm}^3\:\text{s}^{-1}$ at the time of freeze-out. This can e.g. be achieved through a subdominant wino admixture in $\tilde{\chi}^0_1$ which induces annihilation into $W$ bosons. Alternatively, a thermal cross section can be realised by the annihilation of $\tilde{\chi}^0_1$ into bottom quarks or gluons through the MSSM admixture of $A_1$.

\begin{figure}[hbt]
\centering
\includegraphics[width=12cm]{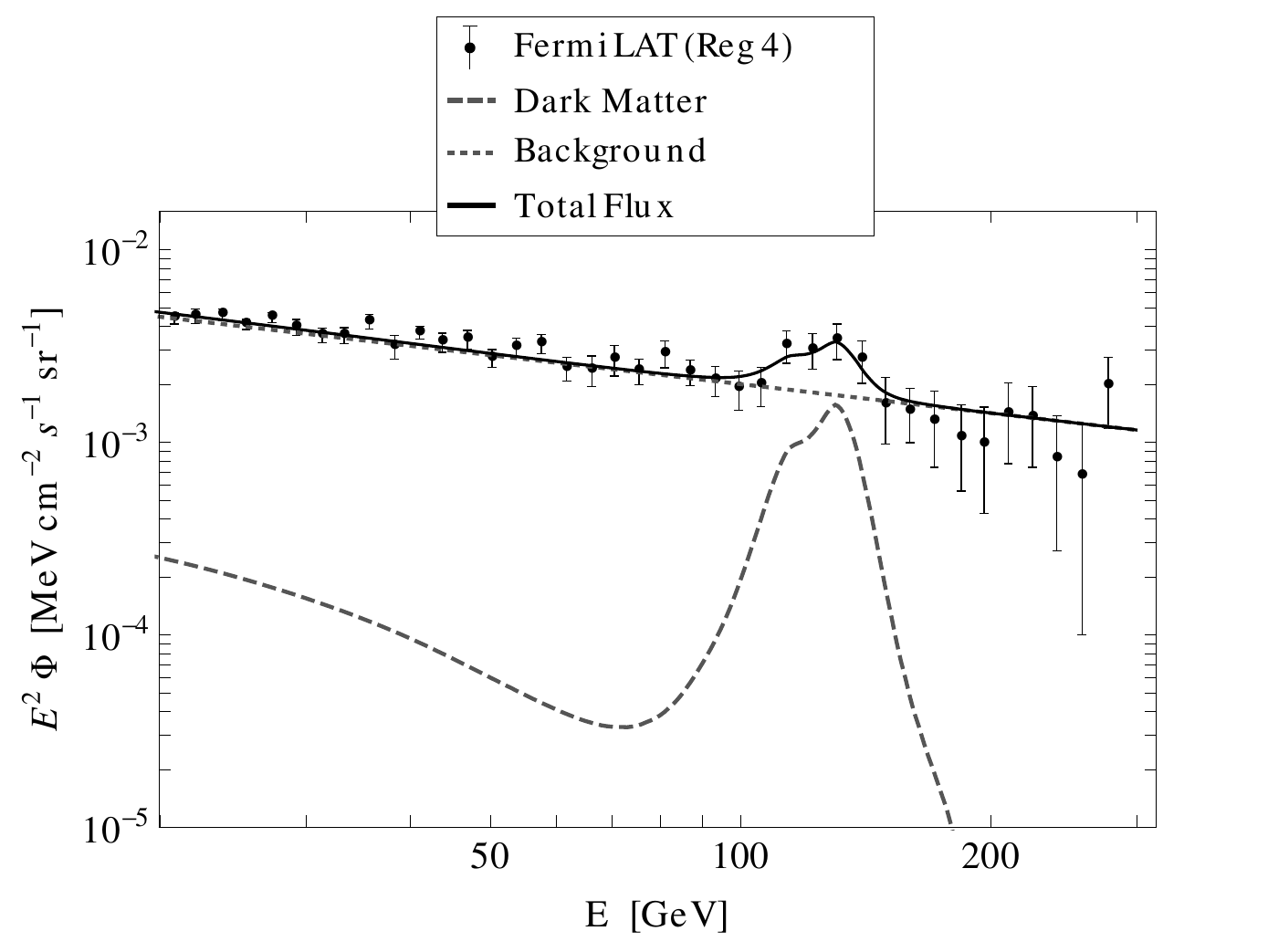}
\caption{$\gamma$ ray flux for the benchmark scenario of table~\ref{tab:bp}. The Fermi data are taken from~\cite{Weniger:2012tx} (region 4, source). The dark matter induced flux comprises the lines from annihilation into $\gamma\gamma$ and $\gamma Z$ as well as the continuum photons from the fragmentation and decay of the accompanying annihilation products. The background is modelled with a power law.}
\label{fig:linefit}
\end{figure}

In figure~\ref{fig:linefit} we present the $\gamma$ ray flux for a specific benchmark choice of the GNMSSM parameters which can be found in table~\ref{tab:bp}. The benchmark scenario will be discussed in more detail in section~\ref{sec:spheno}. For the dark matter density distribution we assume an Einasto profile with the parameters as given in~\cite{Weniger:2012tx}. The Fermi data are taken from region 4 (source) of the same reference. The $\gamma$ ray lines from dark matter annihilation into $\gamma\gamma$ and $Z\gamma$ have been convoluted with the Fermi energy resolution as extracted from table~3 in~\cite{Weniger:2012tx}. The $\gamma$ ray background is modeled with a featureless power law with index $-2.5$. 

It can be seen that the double line at $E=130\gev$ and $E=114\gev$ (from annihilation into $\gamma\gamma$ and $Z\gamma$ respectively) gives a very good fit to the data. The continuum $\gamma$s which -- in the benchmark scenario -- mainly arise from the $WW$-channel are sufficiently suppressed, i.e. they are nicely consistent with the low energy data.

\section{Direct and indirect detection constraints}
\label{sec:dd}

Let us now briefly discuss constraints on the model which arise from direct dark matter searches. The spin-independent cross section of the lightest neutralino with nucleons $\sigma_n^\text{SI}$ is typically dominated by exchange of the light scalar Higgs $h_1$. It can be written in the form\footnote{We neglect the small differences between proton and neutron.}
\begin{equation}
 \sigma_n^\text{SI} \simeq \frac{4 \, m_n^4}{\pi}  \,|f_q|^2\,\left(f^n_u + f^n_d + f^n_s + \frac{6}{27} \,f^n_G \right)^2\;,
\end{equation}
where $f^n_u$, $f^n_d$, $f^n_s$ and $f^n_G$ denote the up-, down-, strange-quark and gluon contributions to the nucleon mass $m_n$ which we take from~\cite{Belanger:2008sj}. For simplicity, we have applied the decoupling limit on the MSSM Higgs fields such that the effective neutralino quark coupling divided by the quark mass $f_q$ is universal among the quark families. The latter takes the form (see e.g.~\cite{Kappl:2010qx})
\begin{equation}\label{eq:quarkcoupl}
 |f_q|^2 = \frac{G_F}{2\,\sqrt{2}\,m_{h_1}^4}\,\left|g_{\tilde{\chi}^0_1  \tilde{\chi}^0_1  h_1}\right|^2\,\left(1-{Z^H_{13}}^2\right)\,,
\end{equation}
where $G_F$ is the Fermi constant and ${Z^H_{13}}^2$ the singlet fraction of the light Higgs $h_1$ (see appendix~\ref{sec:massmatrices}). 
The neutralino Higgs coupling is given as $g_{\tilde{\chi}^0_1  \tilde{\chi}^0_1  h_1}=g_{\tilde{\chi}^0_1  \tilde{\chi}^0_1  h_1}^R+g_{\tilde{\chi}^0_1  \tilde{\chi}^0_1  h_1}^L$ with the left and right couplings being defined in appendix~\ref{sec:couplings}. Note that they depend strongly on the composition of the lightest neutralino.

The relevant limit on the cross section $\sigma_n^\text{SI}$ is set by the XENON100 experiment, which for a neutralino mass $m_{\tilde{\chi}^0_1 }\simeq 130\gev$ corresponds to~\cite{Aprile:2012nq}
\begin{equation}
 \sigma_n^\text{SI} \leq 3\cdot 10^{-45}\:\text{cm}^2\;.
\end{equation}
By use of~\eqref{eq:quarkcoupl} we can translate this into a limit on the coupling $g_{\tilde{\chi}^0_1  \tilde{\chi}^0_1  h_1}^{L,R}$. Assuming that $h_1$ is dominantly SM like, we estimate $|g_{\tilde{\chi}^0_1  \tilde{\chi}^0_1  h_1}| \lesssim 0.05$.\footnote{Note, however, that the uncertainties in the nucleon composition may affect this constraint.}

As described in section~\ref{sec:continuum}, we are mainly interested in the case where $\tilde{\chi}^0_1$ is singlino-like such that the production of continuum $\gamma$s is suppressed. However, in order to enhance annihilation of $\tilde{\chi}^0_1$ into photon pairs, it is favourable to have light (charged) higgsinos. Thus $\tilde{\chi}^0_1$ always contains a non-negligible higgsino admixture. In this case $g_{\tilde{\chi}^0_1  \tilde{\chi}^0_1  h_1}$ typically receives comparable contributions through the $\lambda$ as well as the $\kappa$ coupling bearing also the possibility of (partial) cancellations. Still, we find that for $\lambda\sim1$ the higgsino fraction of ${\tilde{\chi}}^0_1$ should not exceed $\sim10\%$ in order to satisfy the XENON bound.

Further constraints on the model arise from the neutrino searches by Super-Kamio-kande \cite{Tanaka:2011uf,Kappl:2011kz} and IceCube~\cite{IceCube:2011aj} which aim to detect the annihilation of dark matter particles in the sun. They provide especially strong bounds on the spin-dependent cross section $\sigma_p^\text{SD}$ of WIMPs with protons. 
Assuming $m_{{\tilde{\chi}}^0_1}\simeq 130\gev$, the relevant upper limit reads $\sigma_p^\text{SD} = 4 \cdot 10^{-40}\:\text{cm}^2$ \cite{Tanaka:2011uf} if the LSPs dominantly annihilate into $W$ bosons. The constraint, however, gets significantly weaker for annihilation channels which induce a softer neutrino spectrum. In the considered region of parameter space, the leading spin-dependent WIMP proton cross section $\sigma_p^\text{SD}$ arises from $Z$ exchange. It scales with the higgsino components of ${\tilde{\chi}}^0_1$, more specifically $\sigma_p^\text{SD} \propto |N_{13}|^2-|N_{14}|^2$ (cf.~appendix~\ref{sec:couplings}). While this cross-section can be sizeable, there generically occur cancellations due to $|N_{13}|\sim |N_{14}|$. The most dangerous situation arises if ${\tilde{\chi}}^0_1$ is strongly mixed among the different states as in this case $N_{13}$ and $N_{14}$ typically get split. Then, a $\sigma_p^\text{SD}$ close to the current experimental bounds may be generated. 

We find that the remaining constraints from indirect dark matter detection, arising e.g.\ from $\gamma$ ray searches in the Milky Way satellite galaxies by Fermi~\cite{Ackermann:2011wa} or antiproton searches by PAMELA~\cite{Adriani:2010rc,Donato:2008jk} and BESS-Polar~II~\cite{Abe:2011nx,Kappl:2011jw}, are in general weaker than those from the continuum $\gamma$s studied previously. To illustrate that in the GNMSSM, the Fermi line can be explained while all direct and indirect detection constraints are satisfied, we provide an explicit example in the next section.

\section{A benchmark scenario}
\label{sec:spheno}

After we have gained some analytical understanding, we now turn to a full-fledged numerical analysis. For this purpose we use the \SPheno version \cite{Porod:2003um,Porod:2011nf} for the GNMSSM created by \SARAH \cite{Staub:2008uz,Staub:2009bi,Staub:2010jh,Staub:2012pb} which has been presented in \cite{Ross:2012nr}. This version performs a complete one-loop calculation of all SUSY and Higgs masses \cite{Pierce:1996zz,Staub:2010ty} and includes the dominant two-loop corrections for the scalar Higgs masses \cite{Dedes:2003km,Dedes:2002dy,Brignole:2002bz,Brignole:2001jy}. In addition, it calculates the decay widths and branching ratios of all SUSY and Higgs particles. In the Higgs sector the decays are calculated with the following precision: the channels with two SUSY particles, SM leptons or SM vector bosons in the final state are calculated at tree level. In contrast, for quark final states the dominant one-loop QCD corrections due to gluons are included \cite{Spira:1995rr}. For the decays into two photons and two gluons induced at one-loop level all possible leading order contributions are included. In addition, for the CP even Higgs also the dominant NLO QCD corrections are added \cite{Spira:1995rr}. 
Furthermore, this \SPheno version also includes routines to calculate $b\to s\gamma$, $\delta\rho$ and $g-2$ which have been used to check possible constraints from these observables. These calculations are performed in the GNMSSM with the same precision as described in Ref.~\cite{Porod:2011nf} for the MSSM. 

For the calculation of the relic density of the lightest neutralino as well as to obtain $\langle \sigma v\rangle_{\gamma \gamma}$ we have used {\tt MicrOmegas} \cite{Belanger:2006is,Belanger:2007zz,Belanger:2010pz}. For this purpose we created model files for {\tt CalcHep} \cite{Pukhov:2004ca} with \SARAH. These model files include optionally also the effective interactions $h_i\gamma\gamma$ and $A^h_i\gamma\gamma$. The numerical values for these operators as well as of all other parameters are read from the spectrum file written by \SPheno using the {\tt SLHA+} functionality of {\tt CalcHep} \cite{Belanger:2010st}. We also used {\tt MicrOmegas} to calculate the continuous $\gamma$ spectrum which has already been discussed in section~\ref{sec:continuum}. 
\begin{table}[!h!]
\centering
 \begin{tabular}{|p{5cm} c || p{5cm} c|}
\hline
\hline
\multicolumn{4}{|c|}{Input} \\
\hline
$\tan \beta$                                &1.2               &  $v_s$       [GeV]                           &-4.0            \\
$\lambda$                                   &0.74             &   $A_\lambda$ [GeV]                           &$0$  \\
$\kappa$                                    &1.4              & $A_\kappa$  [GeV]                           &$0$ \\
$\mu_s$     [GeV]                           &  103.0         &$b_s~[\text{GeV}^2]$  & $3.356\cdot 10^5$\\
$\mu$  [GeV]                                &  280.0           &$b\mu~[\text{GeV}^2]$                        & $2.4\cdot 10^5$\\
$M_1$  [GeV]                      &       1500.0          &                $M_2$ [GeV]& 193.0 \\
$M_3$  [GeV]                      &        1500.0         &            $m_{scalar}$ [GeV]& 1500.0 \\
$A_{top} Y_{top}$ [GeV]                     & 1500.0             &$\xi_S$ [GeV$^3$]& 0.0\\
\hline
\hline
\multicolumn{4}{|c|}{CP even Higgs sector} \\
\hline
$m_{h_1}$ [GeV] &  125.7  &  down fraction $h_1$ & 41.5\%   \\
$m_{h_2}$ [GeV] &  690.1  &  up fraction $h_1$ &   57.8\% \\
$m_{h_3}$ [GeV] &  786.8  &  singlet fraction $h_1$ &   0.7\% \\
\hline
\hline
\multicolumn{4}{|c|}{CP odd Higgs sector} \\
\hline
$m_{A_1}$ [GeV] &  247.5  &  singlet fraction $A_1$ & 99.9\%    \\
$m_{A_2}$ [GeV] &  691.9  &  up and down fraction $A_1$ &   0.1\% \\
\hline 
\hline  
\multicolumn{4}{|c|}{Neutralino sector} \\
\hline
$\tilde{\chi}^0_1$ [GeV] &  130.0  &  bino fraction $\tilde{\chi}^0_1$ & $<$0.1\%    \\
$\tilde{\chi}^0_2$ [GeV] &  156.4  &  wino fraction $\tilde{\chi}^0_1$ &  5.1\%  \\
$\tilde{\chi}^0_3$ [GeV] &  316.2  &  down-higgsino fraction $\tilde{\chi}^0_1$ & 0.3\%   \\
$\tilde{\chi}^0_4$ [GeV] &  331.6  &  up-higgsino fraction $\tilde{\chi}^0_1$ & 10.0\%   \\
$\tilde{\chi}^0_5$ [GeV] &  1497.4  &  singlet fraction $\tilde{\chi}^0_1$ & 84.5\%   \\
\hline
\hline  
\multicolumn{4}{|c|}{Chargino sector} \\
\hline
$\tilde{\chi}^+_1$ [GeV] &  154.8  &  wino fraction $\tilde{\chi}^+_1$ & 70.6\%   \\
$\tilde{\chi}^+_2$ [GeV] &  332.6  &  higgsino fraction $\tilde{\chi}^+_1$ & 29.4\%   \\
\hline
\hline
\multicolumn{4}{|c|}{Electroweak observables} \\
\hline               
$R_{\gamma\gamma}$& 1.2  &$R_{b \bar{b}}$& 1.0\\
$R_{Z Z}$& 1.0  &$R_{\tau \bar{\tau}}$& 1.0\\
$\text{Br}(b \rightarrow s \gamma)$         & $3.4 \cdot 10^{-4}$ & $\text{Br}(B_s \rightarrow \mu\mu)$ & $3.7 \cdot 10^{-9}$ \\
 $\Delta a_\mu$                              & $-1.2 \cdot 10^{-11}$ & $\delta\rho$                                &  $4.5 \cdot 10^{-5}$  \\
\hline
\hline
\multicolumn{4}{|c|}{Dark matter} \\
\hline               
$\Omega h^2$                                & 0.1  & $X_{FO}$  & 24.9 \\
$\sigma_p^\text{SI}\, [\text{cm}^2]$           & $2.2 \cdot 10^{-45}$   &$ \sigma_p^\text{SD}\, [\text{cm}^2]$                           & $3.8 \cdot 10^{-40}$   \\ 
$\langle \sigma v\rangle_{\gamma \gamma}\, [\text{cm}^3/\text{s}]$ &	$0.83 \cdot 10^{-27}$	& $\langle \sigma v\rangle_{\gamma Z}\,  [\text{cm}^3/\text{s}]$ & $0.79 \cdot 10^{-27}$		\\
\hline
\hline
 \end{tabular}
\caption{Benchmark point for the GNMSSM. $R_{XY}$ denotes the production cross section of $pp\to h \to XY$ normalised to the SM expectations (based on the values of the CERN yellow pages \cite{CernYellow}). }
\label{tab:bp}
\end{table}
\begin{figure}[t]
\centering 
\begin{minipage}{\linewidth}
\includegraphics[height=0.3\linewidth]{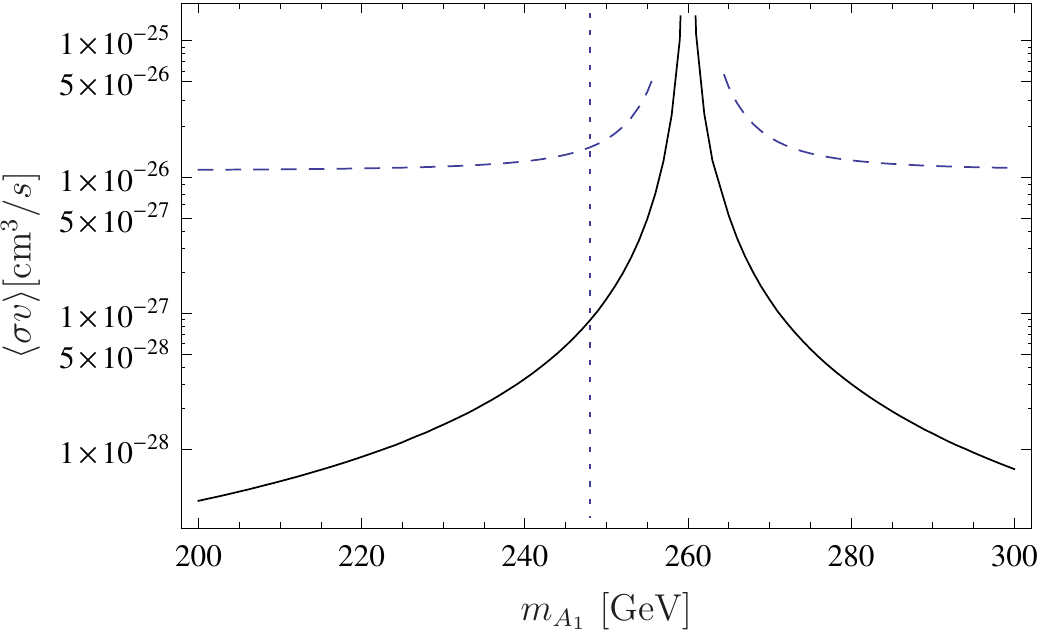}
\hfill
\includegraphics[height=0.3\linewidth]{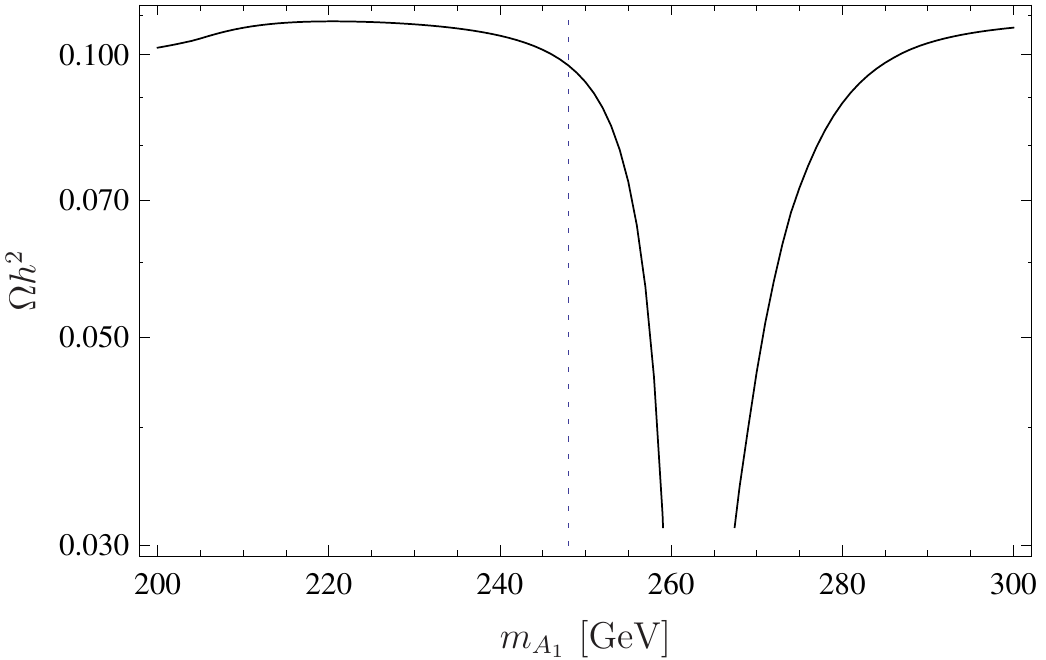}
\end{minipage}
\caption{Left: The present day diphoton rate $\langle \sigma v\rangle_{\gamma \gamma}$ (solid line) and the total cross section $\langle \sigma v\rangle$ (dashed line) as function of the light pseudoscalar mass $m_{A_1}$. Right: dependence of the dark matter relic density $\Omega h^2$ on the pseudoscalar mass. 
The other parameters are those of the point given in table~\ref{tab:bp}. The dotted, vertical line indicates the pseudoscalar mass for our benchmark point.}
\label{fig:gamma}
\end{figure}

In table~\ref{tab:bp} we show a benchmark point with all the desired features: the light Higgs mass is close to 125~GeV and the branching ratios into two photons is enhanced by 20\% because of the chargino loop contributions. The LSP is the lightest neutralino which is mostly singlino-like and has a mass of 130~GeV. A mostly singlino LSP can be achieved by appropriate values for $M_1$, $M_2$ and $\mu_\text{eff}$. While $M_1$ plays only a subleading role as long as the bino is heavier than the singlino, the choice of $\mu_\text{eff}$ has to be done more carefully: on the one hand it should not be too small in order to suppress the mixing between the singlino and higgsino, because a sizeable higgsino fraction is often in conflict with direct detection measurements. On the other hand, light charginos with a sizeable higgsino fraction are needed in order to enhance the loop contributions to $h\to\gamma\gamma$ and $\langle \sigma v\rangle_{\gamma\gamma}$. For the benchmark point a light chargino has been realized by a comparably small value of $M_2$, which however leads to a large mixing in the chargino sector. Note that the benchmark scenario is consistent with direct chargino searches at the LHC as the latter have only gained sensitivity to spectra with $m_{\tilde{\chi}^0_1} \lesssim 100\gev$~\cite{ATLASCHARGINO,CMSCHARGINO}.\footnote{The sensitivity of chargino searches increases if there exists a slepton with mass between $m_{\tilde{\chi}^+_1}$ and $m_{\tilde{\chi}^0_1}$. This is, however, not the case in our benchmark scenario.}
The correct relic density is obtained mainly via annihilation into $W^+W^-$, while today's annihilation into photons mainly proceeds via the pseudoscalar exchange. This pseudoscalar is nearly a pure singlet and it is not necessary to be very close to the resonance: even with a mass more than 10~GeV away from the resonance the diphoton rate is enhanced to a level sufficient to explain the tentative Fermi line. 
This is a big improvement in comparison to the NMSSM where one has to be usually very close to the resonance: for cases with the pseudoscalar component of the singlet in the correct mass range, the singlino fraction of the LSP is very small and the coupling between both is highly suppressed. This is not only a drawback of the NMSSM with respect to the needed fine-tuning, but, what is even more important, this scenario is also under big pressure from direct detections bounds. The dependence of $\langle \sigma v\rangle$ on $m_{A_1}$ is shown in figure~\ref{fig:gamma} (left), with all other parameters as given in table~\ref{tab:bp}. In the right panel we also show the relic density $\Omega h^2$, which in the region of interest is hardly effected by the pseudoscalar resonance.

\begin{figure}[hbt]
\centering
\includegraphics[width=10cm]{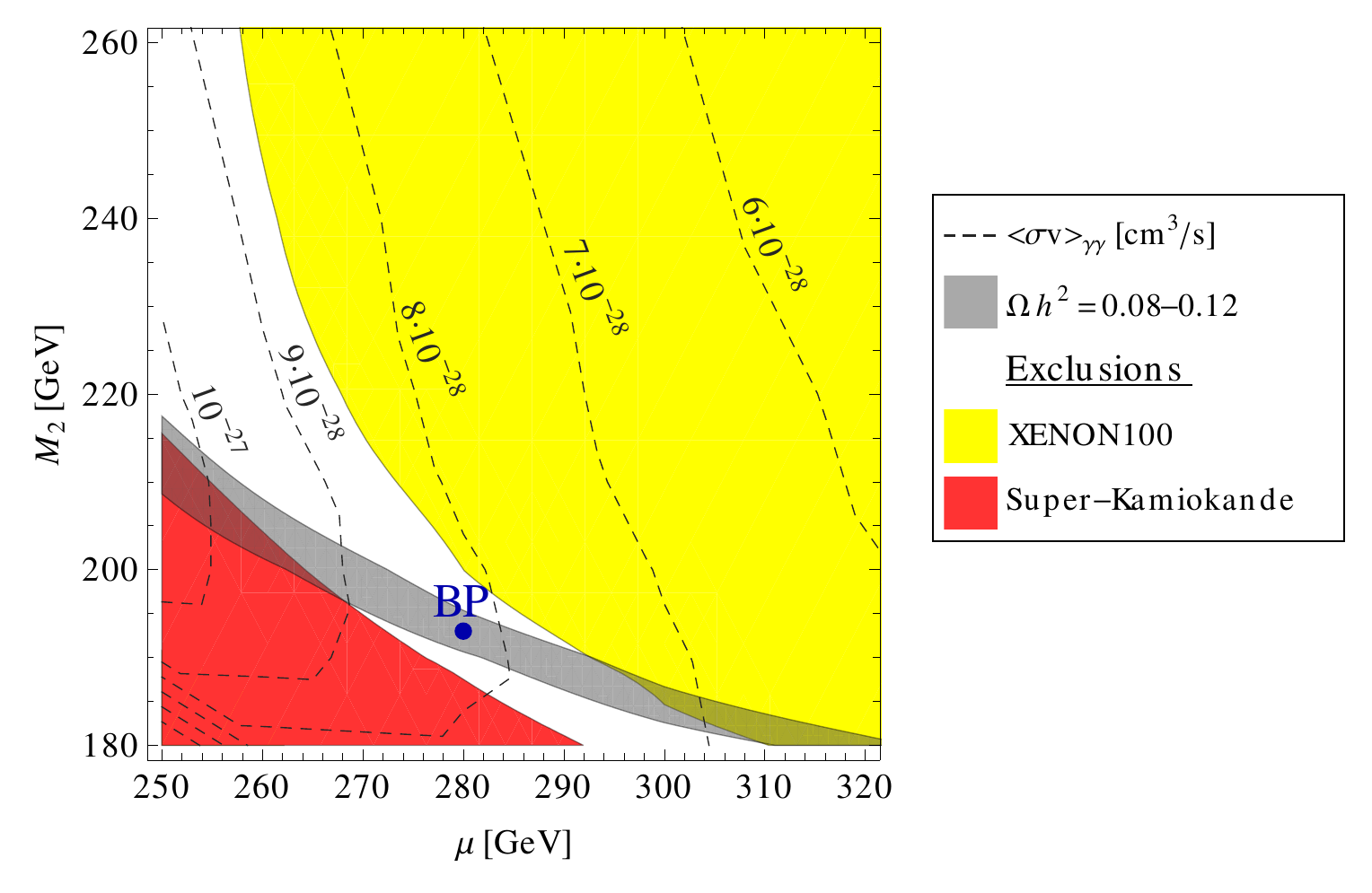}
\caption{Parameter Scan in the GNMSSM (see text) with contours referring to the value of $\langle\sigma v\rangle_{\gamma\gamma}$. The benchmark scenario of table~\ref{tab:bp} is indicated by the blue dot. In the gray band, the thermal relic density of the lightest neutralino is consistent with the observed dark matter density. The yellow region is exluded by the XENON100 direct dark matter search, the red region is excluded by the Super-Kamiokande limit on the spin-dependent cross section of $\tilde{\chi}^0_1$.}
\label{fig:scan}
\end{figure}

In order to illustrate the constraints on the parameter space of the GNMSSM which arise from direct and indirect detection, we show the results of a scan over $\mu$ and $M_2$ in figure~\ref{fig:scan}. For each combination $(\mu,M_2)$ we adjusted the parameters $\mu_s$ and $b_s$ such that $m_{\tilde{\chi}^0_1}$ and $m_{A_1}$ remain fixed at their values from table~\ref{tab:bp}. The remaining parameters were taken from the same table. 

It can be seen that, while $\tilde{\chi}^0_1$ is dominated by its singlino component in the depicted parameter space, a rather low $M_2$ is favorable in order to satisfy the direct detection constraints. This is because the Higgs, which mediates the spin-independent interaction of $\tilde{\chi}^0_1$ with nucleons, mainly couples to the higgsino and singlino components of $\tilde{\chi}^0_1$ due to the relatively large $\lambda$ and $\kappa$. An additional wino admixture tends to reduce $\sigma_p^\text{SI}$. At the same time, the wino component of $\tilde{\chi}^0_1$ drives its annihilation into $W$ pairs allowing for a thermal neutralino abundance which agrees with the dark matter abundance (gray band in figure~\ref{fig:scan}). On the other hand, a light wino induces a splitting between the higgsino components of the lightest neutralino, $N_{13}$ and $N_{14}$. This, in turn, enhances the spin-dependent cross section $\sigma_p^\text{SD}$ (see section~\ref{sec:dd}) resulting in the Super-Kamiokande excluded region at low $M_2$ and $\mu$.\footnote{To obtain the Super-Kamiokande limit on the spin-dependent cross section, we have weighted $\sigma_p^\text{SI}$ by the fractional annihilation of $\tilde{\chi}^0_1$ into channels which induce hard neutrinos.} The annihilation of $\tilde{\chi}^0_1$ into $\gamma$-pairs is mediated by higgsinos in the loop, while winos do virtually not contribute due to the singlet nature of $A_1$. Therefore, $\langle \sigma v\rangle_{\gamma \gamma}$ grows with decreasing higgsino mass. Note, finally, that the direct and indirect detection constraints get weaker for smaller couplings $\lambda$ and $\kappa$. In this case, however, one would need $m_{A_1}$ closer to the resonance in order to keep $\langle \sigma v\rangle_{\gamma \gamma}$ large.

\section{Summary}
\label{sec:conclusions}

In this article we have shown that within the GNMSSM, a generalised version of the NMSSM, the experimental hints beyond the SM from the Fermi LAT telescope as well as from the LHC can be explained simultaneously while being consistent with all experimental constraints. As in the NMSSM the superpotential term $\lambda\,S H_u H_d$ plays a crucial role in this scheme: it drives the annihilation of the lightest neutralino into photons, induces new tree-level contributions to the mass of the light Higgs boson $h$ and enhances the partial width of the decay $h\rightarrow \gamma\gamma$. 
To obtain a large enough annihilation cross section into photons, a very mild tuning of the pseudoscalar Higgs mass is required in the GNMSSM: it is sufficient to be within $\sim 20 \gev$ from the resonance. This is in contrast to the NMSSM, where the tuning is very substantial, because unlike in the GNMSSM a mainly singlino-like LSP is not possible, leading to much smaller effective couplings. Hence while the coupling $\lambda$ is also present in the NMSSM, it is the additional flexibility in the mass spectrum of the GNMSSM which allows to simultaneously explain the Fermi $\gamma$ ray line, ameliorate the little hierarchy problem, and explain a moderate excess in the $\gamma\gamma$-channel which is indicated by the Higgs searches of CMS and ATLAS. 

To substantiate our claim we performed a thorough numerical analysis and presented an example point which features a $\sim 130 \gev$ lightest neutralino with an annihilation cross section into photons consistent with the indication from the Fermi satellite with simultaneously the right relic abundance, a continuum photon spectrum consistent with observation, direct detection cross section below the experimental limits, electroweak observables consistent with experiment and a $125 \gev$ light Higgs boson with a slightly enhanced $h \rightarrow \gamma \gamma$ rate. 

\section*{Acknowledgements}
We would like to thank Mathias Garny, Chris McCabe, and Graham G.~Ross for useful discussions. 
\appendix

\section{Mass matrices}
\label{sec:massmatrices}
If we decompose the complex Higgs fields and singlet after EWSB as
\begin{align} 
H_d^0 = & \, \frac{1}{\sqrt{2}} \left(\phi_{d}  +  v_d  + i \sigma_{d} \right)\\ 
H_u^0 = & \, \frac{1}{\sqrt{2}} \left(\phi_{u}  +  v_u  + i  \sigma_{u} \right)\\ 
S = & \, \frac{1}{\sqrt{2}} \left(\phi_s  + v_s  + i \sigma_s \right)
\end{align}
the mass matrices in the neutral Higgs sector read:
\begin{itemize}
\item Scalar Higgs. Basis: $\left(\phi_{d}, \phi_{u}, \phi_s\right)$ \\
\begin{align} 
 \label{eq:scalarMM}
m_{dd} &= \frac{1}{8} \Big(4 \Big(2 \mu  + \sqrt{2} v_s \lambda \Big)\mu^*  + 4 \Big(\sqrt{2} v_s \mu  + \Big(v_{s}^{2} + v_{u}^{2}\Big)\lambda \Big)\lambda^*  \nonumber \\ & \hspace{1.5cm} + 8 m_{H_d}^2  + \Big(g_{1}^{2} + g_{2}^{2}\Big)\Big(3 v_{d}^{2}  - v_{u}^{2} \Big)\Big)\\ 
m_{ud} &= \frac{1}{4} \Big(-2 v_{s}^{2} {\Re\Big(\lambda \kappa^* \Big)}  -4 {\Re\Big(b \mu\Big)}    + 4 v_d v_u |\lambda|^2  \nonumber \\ & \hspace{1.5cm}  - \Big(g_{1}^{2} + g_{2}^{2}\Big)v_d v_u  - \sqrt{2} v_s \Big(2 {\Re\Big(\lambda \mu_S^* \Big)}  + 2 {\Re\Big(\lambda A_{\lambda}\Big)} \Big)\Big)\\ 
m_{uu} &= \frac{1}{8} \Big(4 \Big(2 \mu  + \sqrt{2} v_s \lambda \Big)\mu^*  + 4 \Big(\sqrt{2} v_s \mu  + \Big(v_{d}^{2} + v_{s}^{2}\Big)\lambda \Big)\lambda^*  + 8 m_{H_u}^2  \nonumber \\ & \hspace{1.5cm}  - \Big(g_{1}^{2} + g_{2}^{2}\Big)\Big(-3 v_{u}^{2}  + v_{d}^{2}\Big)\Big)\\ 
m_{sd} &= -\frac{1}{4} v_u \Big(4 v_s {\Re\Big(\lambda \kappa^* \Big)}  + \sqrt{2} \Big(2 {\Re\Big(\lambda A_{\lambda}\Big)}  + \mu_S \lambda^* \Big) + \sqrt{2} \lambda \mu_S^* \Big) \nonumber \\ &  \hspace{1.5cm} + \frac{1}{\sqrt{2}} v_d \lambda \mu^*  + v_d \Big(\frac{1}{\sqrt{2}} \mu  + v_s \lambda \Big)\lambda^* \\ 
m_{su} &= -\frac{1}{4} v_d \Big(4 v_s {\Re\Big(\lambda \kappa^* \Big)}  + \sqrt{2} \Big(2 {\Re\Big(\lambda A_{\lambda}\Big)}  + \mu_S \lambda^* \Big) + \sqrt{2} \lambda \mu_S^* \Big) + \frac{1}{\sqrt{2}} v_u \lambda \mu^*  \nonumber \\ &  \hspace{1.5cm} + \Big(\frac{1}{\sqrt{2}} v_u \mu  + v_s v_u \lambda \Big)\lambda^* \\ 
m_{ss} &= \frac{1}{2} \Big(6 v_{s}^{2} |\kappa|^2 +v_{d}^{2} |\lambda|^2 +v_{u}^{2} |\lambda|^2  +2 \Big(m_S^2 + |\mu_S|^2 + {\Re\Big(b_S\Big)}\Big)\nonumber \\ &  \hspace{1.5cm} -2 v_d v_u {\Re\Big(\lambda \kappa^* \Big)} +\sqrt{2} v_s \Big(2 {\Re\Big(\kappa A_{\kappa}\Big)}  + 6 {\Re\Big(\kappa \mu_S^* \Big)} \Big)\Big)
\end{align} 
\item Pseudoscalar Higgs. Basis: $\left(\sigma_{d}, \sigma_{u}, \sigma_s\right)$ \\
 \begin{align} 
m_{dd} &= \frac{1}{8} \Big(4 \Big(2 \mu  + \sqrt{2} v_s \lambda \Big)\mu^*  + 4 \Big(\sqrt{2} v_s \mu  + \Big(v_{s}^{2} + v_{u}^{2}\Big)\lambda \Big)\lambda^*  \\ & \hspace{1.5cm}  + 8 m_{H_d}^2  + \Big(g_{1}^{2} + g_{2}^{2}\Big)\Big(- v_{u}^{2}  + v_{d}^{2}\Big)\Big)\\ 
m_{ud} &= \frac{1}{4} \Big(4 {\Re\Big(b \mu\Big)}  + v_s \Big(2 v_s {\Re\Big(\lambda \kappa^* \Big)}  + \sqrt{2} \Big(2 {\Re\Big(\lambda \mu_S^* \Big)}  + 2 {\Re\Big(\lambda A_{\lambda}\Big)} \Big)\Big)\Big)\\ 
m_{uu} &= \frac{1}{8} \Big(4 \Big(2 \mu  + \sqrt{2} v_s \lambda \Big)\mu^*  + 4 \Big(\sqrt{2} v_s \mu  + \Big(v_{d}^{2} + v_{s}^{2}\Big)\lambda \Big)\lambda^*  \\ & \hspace{1.5cm}  + 8 m_{H_u}^2  - \Big(g_{1}^{2} + g_{2}^{2}\Big)\Big(- v_{u}^{2}  + v_{d}^{2}\Big)\Big)\\ 
m_{sd} &= -\frac{1}{4} v_u \Big(4 v_s {\Re\Big(\lambda \kappa^* \Big)}  + \sqrt{2} \Big(-2 {\Re\Big(\lambda A_{\lambda}\Big)}  + \mu_S \lambda^* \Big) + \sqrt{2} \lambda \mu_S^* \Big)\\ 
m_{su} &= -\frac{1}{4} v_d \Big(4 v_s {\Re\Big(\lambda \kappa^* \Big)}  + \sqrt{2} \Big(-2 {\Re\Big(\lambda A_{\lambda}\Big)}  + \mu_S \lambda^* \Big) + \sqrt{2} \lambda \mu_S^* \Big)\\ 
m_{ss} &= \frac{1}{2} \Big(2 v_{s}^{2} |\kappa|^2 +v_{d}^{2} |\lambda|^2 +v_{u}^{2} |\lambda|^2 +2 \Big(  - {\Re\Big(b_S\Big)}  + m_S^2 \nonumber \\ 
 & \hspace{1.5cm} + |\mu_S|^2\Big)+2 v_d v_u {\Re\Big(\lambda \kappa^* \Big)} +\sqrt{2} v_s \Big(2 {\Re\Big(\kappa \mu_S^* \Big)}  -2 {\Re\Big(\kappa A_{\kappa}\Big)} \Big)\Big)
\end{align} 
\end{itemize}
After eliminating the Goldstone mode, the mass matrix can be written as
\begin{align}
M_{1,1} &= \left(v_s (\sqrt{2}(A_\lambda + \mu_S) + v_s \kappa)\lambda + 2 b\mu\right)/\sin(2\beta) \\
M_{1,2} &= \frac{v}{\sqrt{2}} \lambda (A_\lambda - \sqrt{2} v_s \kappa - \mu_s) \\
M_{2,2} &= -2 b_s - \frac{3}{\sqrt{2}} A_\kappa v_s \kappa - (v^2 \lambda \mu)/(\sqrt{2}v_s) - v_s \kappa \mu_s/\sqrt{2} - \sqrt{2}\xi_s/v_s \nonumber \\
            &+ ((A_\lambda  +  \mu_s) \lambda v^2  \cos\beta \sin\beta)/(\sqrt{2}v_s)
         +2 v^2 \kappa \lambda \cos\beta \sin\beta 
\end{align}
The mass matrices of the charged Higgs as well as of the neutralinos and charginos are given by
\begin{itemize}
\item Charged Higgs.  Basis: $\left(H_d^-, H_u^{+,*}\right)$\\
\begin{align}
m_{dd} &= \frac{1}{8} \Big(4 \Big(2 \mu  + \sqrt{2} v_s \lambda \Big)\mu^*  + 4 v_s \Big(\sqrt{2} \mu  + v_s \lambda \Big)\lambda^*  + 8 m_{H_d}^2 \nonumber \\ 
 & \hspace{1.5cm}  + g_{1}^{2} \Big(- v_{u}^{2}  + v_{d}^{2}\Big) + g_{2}^{2} \Big(v_{d}^{2} + v_{u}^{2}\Big)\Big)\\ 
m_{ud} &= \frac{1}{4} \Big(2 v_s \Big(\sqrt{2} \Big(\lambda \mu_S^*  + \lambda A_{\lambda}\Big) + v_s \lambda \kappa^* \Big) + 4 b \mu   - v_d v_u \Big(2 |\lambda|^2  - g_{2}^{2} \Big)\Big)\\ 
m_{uu} &= \frac{1}{8} \Big(4 \sqrt{2} v_s \lambda \mu^*  + 4 v_s \Big(\sqrt{2} \mu  + v_s \lambda \Big)\lambda^*  + 8 m_{H_u}^2  \nonumber \\ 
 & \hspace{1.5cm} + 8 |\mu|^2  + \Big(g_{1}^{2} + g_{2}^{2}\Big)v_{u}^{2}  - g_{1}^{2} v_{d}^{2}  + g_{2}^{2} v_{d}^{2} \Big)
\end{align} 
 \item Neutralino. Basis: $\left(\lambda_{\tilde{B}}, \tilde{W}^0, \tilde{H}_d^0, \tilde{H}_u^0, \tilde{S}\right)$\\
 \begin{equation} 
m_{N} = \left( 
\begin{array}{ccccc}
M_1 &0 &-\frac{1}{2} g_1 v_d  &\frac{1}{2} g_1 v_u  &0\\ 
0 &M_2 &\frac{1}{2} g_2 v_d  &-\frac{1}{2} g_2 v_u  &0\\ 
-\frac{1}{2} g_1 v_d  &\frac{1}{2} g_2 v_d  &0 &- \frac{1}{\sqrt{2}} v_s \lambda  - \mu  &- \frac{1}{\sqrt{2}} v_u \lambda \\ 
\frac{1}{2} g_1 v_u  &-\frac{1}{2} g_2 v_u  &- \frac{1}{\sqrt{2}} v_s \lambda  - \mu  &0 &- \frac{1}{\sqrt{2}} v_d \lambda \\ 
0 &0 &- \frac{1}{\sqrt{2}} v_u \lambda  &- \frac{1}{\sqrt{2}} v_d \lambda  &\sqrt{2} v_s \kappa  + \mu_S\end{array} 
\right) 
 \end{equation} 
 \item Chargino. Basis: $\left(\tilde{W}^-, \tilde{H}_d^-\right), \left(\tilde{W}^+, \tilde{H}_u^+\right)$
 \begin{equation} 
 \label{eq:CharginoMM}
m_{Ch} = \left( 
\begin{array}{cc}
M_2 &\frac{1}{\sqrt{2}} g_2 v_u \\ 
\frac{1}{\sqrt{2}} g_2 v_d  &\frac{1}{\sqrt{2}} v_s \lambda  + \mu\end{array} 
\right) 
 \end{equation} 
\end{itemize}

\section{Couplings}\label{sec:couplings}
\begin{itemize}
 \item $\tilde{\chi}^0_i \tilde{\chi}^0_j A^h_k$
 \begin{align} 
 g^L_{\tilde{\chi}^0_i \tilde{\chi}^0_j A^h_k} = &\frac{1}{2} \Big(- g_2 N^*_{i 2} N^*_{j 3} Z_{{k 1}}^{A} - \sqrt{2} \lambda N^*_{i 5} N^*_{j 4} Z_{{k 1}}^{A} - \sqrt{2} \lambda N^*_{i 4} N^*_{j 5} Z_{{k 1}}^{A} - g_1 N^*_{i 4} N^*_{j 1} Z_{{k 2}}^{A} \nonumber \\ 
 &+g_2 N^*_{i 4} N^*_{j 2} Z_{{k 2}}^{A} - \sqrt{2} \lambda N^*_{i 5} N^*_{j 3} Z_{{k 2}}^{A} +g_2 N^*_{i 2} N^*_{j 4} Z_{{k 2}}^{A} \nonumber \\ 
 &+g_1 N^*_{i 1} \Big(N^*_{j 3} Z_{{k 1}}^{A}  - N^*_{j 4} Z_{{k 2}}^{A} \Big)- \sqrt{2} \lambda N^*_{i 4} N^*_{j 3} Z_{{k 3}}^{A} +2 \sqrt{2} \kappa N^*_{i 5} N^*_{j 5} Z_{{k 3}}^{A} \nonumber \\ 
 &+N^*_{i 3} \Big(g_1 N^*_{j 1} Z_{{k 1}}^{A}  - g_2 N^*_{j 2} Z_{{k 1}}^{A}  - \sqrt{2} \lambda \Big(N^*_{j 4} Z_{{k 3}}^{A}  + N^*_{j 5} Z_{{k 2}}^{A} \Big)\Big)\Big)\nonumber\\ 
 g^R_{\tilde{\chi}^0_i \tilde{\chi}^0_j A^h_k} = &  -(g^L_{\tilde{\chi}^0_i \tilde{\chi}^0_j A^h_k})^* | (i\leftrightarrow j)
 \end{align} 
 \item $\tilde{\chi}^+_i \tilde{\chi}^-_j A^h_k$
 \begin{align} 
 g^L_{\tilde{\chi}^+_i \tilde{\chi}^-_j A^h_k} = &\frac{1}{\sqrt{2}} \Big(- g_2 U^*_{j 1} V^*_{i 2} Z_{{k 2}}^{A}  + U^*_{j 2} \Big(- g_2 V^*_{i 1} Z_{{k 1}}^{A}  + \lambda V^*_{i 2} Z_{{k 3}}^{A} \Big)\Big)\nonumber\\ 
g^R_{\tilde{\chi}^+_i \tilde{\chi}^-_j A^h_k} = &  -(g^L_{\tilde{\chi}^+_i \tilde{\chi}^-_j A^h_k})^* | (i\leftrightarrow j)  
\end{align} 
\item $\tilde{\chi}^0_i \tilde{\chi}^0_j h_k$
\begin{align} 
 g^L_{\tilde{\chi}^0_i \tilde{\chi}^0_j h_k} = &\frac{i}{2} \Big(- g_2 N^*_{i 2} N^*_{j 3} Z_{{k 1}}^{H} +\sqrt{2} \lambda N^*_{i 5} N^*_{j 4} Z_{{k 1}}^{H} +\sqrt{2} \lambda N^*_{i 4} N^*_{j 5} Z_{{k 1}}^{H} - g_1 N^*_{i 4} N^*_{j 1} Z_{{k 2}}^{H} \nonumber \\ 
 &+g_2 N^*_{i 4} N^*_{j 2} Z_{{k 2}}^{H} +\sqrt{2} \lambda N^*_{i 5} N^*_{j 3} Z_{{k 2}}^{H} +g_2 N^*_{i 2} N^*_{j 4} Z_{{k 2}}^{H} \nonumber \\ 
 &+g_1 N^*_{i 1} \Big(N^*_{j 3} Z_{{k 1}}^{H}  - N^*_{j 4} Z_{{k 2}}^{H} \Big)+\sqrt{2} \lambda N^*_{i 4} N^*_{j 3} Z_{{k 3}}^{H} -2 \sqrt{2} \kappa N^*_{i 5} N^*_{j 5} Z_{{k 3}}^{H} \Big) \nonumber \\
 g^R_{\tilde{\chi}^0_i \tilde{\chi}^0_j h_k} = & - (g^L_{\tilde{\chi}^0_i \tilde{\chi}^0_j h_k})^* | (i\leftrightarrow j)
 \end{align} 
\item $\tilde{\chi}^0_i \tilde{\chi}^0_j Z$
\begin{align} 
 g^L_{\tilde{\chi}^0_i \tilde{\chi}^0_j Z} = &-\frac{i}{2} \Big(g_1 \sin\Theta_W   + g_2 \cos\Theta_W  \Big)\Big(N^*_{j 3} N_{{i 3}}  - N^*_{j 4} N_{{i 4}} \Big)\nonumber\\ 
g^R_{\tilde{\chi}^0_i \tilde{\chi}^0_j Z} =  & (g^L_{\tilde{\chi}^0_i \tilde{\chi}^0_j Z})^* 
\end{align} 
\end{itemize}

\bibliography{NMSSM}
\bibliographystyle{ArXiv}

\end{document}